\documentclass[prb,superscriptaddress,twocolumn]{revtex4-1}
\usepackage{times,amsmath}
\usepackage{epsfig}
\usepackage{color}
\usepackage{longtable}
\usepackage{graphicx}
\usepackage{dcolumn}
\usepackage{bm}
\usepackage{hyperref}
\usepackage{multirow}
\usepackage{amssymb}
\hypersetup{colorlinks=true, citecolor=blue, filecolor=blue,linkcolor=blue, urlcolor=blue}

\begin{document}
\title{Computation and data driven discovery of
topological phononic materials}

\author{Jiangxu Li}
\affiliation{Shenyang National Laboratory for Materials Science,
Institute of Metal Research, \\Chinese Academy of Science, 110016
Shenyang, Liaoning, People Republic of China} \affiliation{School of
Materials Science and Engineering, University of Science and
Technology of China,\\Shenyang 110016, People Republic of China}

\author{Jiaxi Liu}
\affiliation{Shenyang National Laboratory for Materials Science,
Institute of Metal Research, \\Chinese Academy of Science, 110016
Shenyang, Liaoning, People Republic of China} \affiliation{School of
Materials Science and Engineering, University of Science and
Technology of China,\\Shenyang 110016, People Republic of China}

\author{Stanley A. Baronett}
\affiliation{Department of Physics and Astronomy, University of
Nevada, Las Vegas, Nevada 89154, USA}

\author{Ronghan Li}
\affiliation{Shenyang National Laboratory for Materials Science,
Institute of Metal Research, \\Chinese Academy of Science, 110016
Shenyang, Liaoning, People Republic of China} \affiliation{School of
Materials Science and Engineering, University of Science and
Technology of China,\\Shenyang 110016, People Republic of China}

\author{Lei Wang}
\affiliation{Shenyang National Laboratory for Materials Science,
Institute of Metal Research, \\Chinese Academy of Science, 110016
Shenyang, Liaoning, People Republic of China} \affiliation{School of
Materials Science and Engineering, University of Science and
Technology of China,\\Shenyang 110016, People Republic of China}

\author{Qiang Zhu}
\email{qiang.zhu@unv.edu}\affiliation{Department of Physics and
Astronomy, University of Nevada, Las Vegas, Nevada 89154, USA}

\author{Xing-Qiu Chen}
\email{xingqiu.chen@imr.ac.cn} \affiliation{Shenyang National
Laboratory for Materials Science, Institute of Metal Research,
\\Chinese Academy of Science, 110016 Shenyang, Liaoning, People
Republic of China} \affiliation{School of Materials Science and
Engineering, University of Science and Technology of
China,\\Shenyang 110016, People Republic of China}

\date{\today}

\begin{abstract}
The discovery of topological quantum states marks a new chapter in
both condensed matter physics and materials sciences. By analogy to
spin electronic system, topological concepts have been extended into
phonons, boosting the birth of topological phononics (TPs). Here, we
present a high-throughput screening and data-driven approach to
compute and evaluate TPs among over 10,000 materials. We have
clarified 5014 TP materials and classified them into single Weyl,
high degenerate Weyl, and nodal-line (ring) TPs. Among them, three
representative cases of TPs have been discussed in detail.
Furthermore, we suggest 322 TP materials with potential clean
nontrivial surface states, which are favorable for experimental
characterizations. This work significantly increases the current
library of TP materials, which enables an in-depth investigation of
their structure-property relations and opens new avenues for future
device design related to TPs.

\end{abstract}

\maketitle

Over the past decade, topological concepts have made far-reaching
impacts on the theory of electronic band structures in condensed
matters physics and materials sciences~\cite{Kane,Bernevig1757,Fu}.
Thousands of topological electronic materials~\cite{TI1,TI2,TI3}
were theoretically proposed~\cite{ht1,ht2,ht3} and some of them were
experimentally verified, such as, topological
insulators~\cite{TI1,TI2,TI3}, Dirac/Weyl
semimetals~\cite{DSM1,DSM2,WSM1,WSM2,WSM3}, and nodal line
semimetals~\cite{RHLi1-DNL,DNL1,DNL2,DNL3,RHLi2-DNL}. As the
counterpart of electrons, phonons~\cite{phon} are energy quanta of
lattice vibrations. They make crucial contributions to many physical
properties, such as, thermal conductivity, superconductivity and
thermoelectricity as well as specific heat. Similar to topological
electronic nature, the crucial theorems and concepts of topology can
be introduced to the field of phonons, called topological phononics
(TPs)~\cite{TPW2,TPW1,TPW2-2,TPW3,TPW4,TPW5,TPW6,TPNL1,
TPNL2,TPNL3,TPNL4,TPNL5,TPHE,Niu1,Xu1,Xu2,tp1d,exp1,Yongxu2019}. In
particular, TPs in solid materials are also correlated to some
specified atomic lattice vibrations generally within a scale of THz
frequency, thereby providing a rich platform for the investigation
of various quasi-particles related with Bosons.

TPs have been theoretically or experimentally investigated in the
materials. Several theoretical models, including monolayer hexagonal
lattices~\cite{TPHE,Niu1}, Kekul$\acute{e}$ lattice~\cite{Xu1,Xu2}
and one-dimensional (1D) chains~\cite{tp1d}, were discussed. More
recently, a number of real materials were predicted to host the Weyl
TPs~\cite{TPW2,TPW2-2,TPW1,TPW3,TPW4,TPW5,TPW6,TPNL4}, nodal-line
TPs ~\cite{TPNL1,TPNL2,TPNL3,TPNL4,TPNL5} and nodal-ring phonons
\cite{TPNL3}. Single Weyl TPs were predicted in noncentrosymmetric
WC-type materials~\cite{TPW2,TPW2-2}, exhibiting two-fold degenerate
Weyl points with the $\pm1$ topological charges. In FeSi-type
materials, double Weyl TPs were predicted and then experimentally
confirmed~\cite{TPW1,exp1}. In SiO$_2$, the coexisted single/double
Weyl TPs was suggested~\cite{TPW6}. In addition to occupying the
discrete sites in the reciprocal space as Weyl points, these band
crossing points can also continuously form nodal-line (\emph{e.g.},
in MgB$_2$~\cite{TPNL2} and Rb$_2$Sn$_2$O$_3$~\cite{TPNL4}) or
nodal-ring TPs (\emph{e.g.}, in graphene~\cite{TPNL3}, bcc
C$_8$~\cite{TPNL1} and MoB$_2$~\cite{TPNL5}). TPs exhibit the
typical features of bulk-surface (edge) correspondence, which are
rooted in different geometry phases of Hamiltonian. The existence of
multiple critical physical phenomena, such as phononic valley Hall
effect~\cite{Niu1}, phononic quantum anomalous Hall-like effect, and
phononic quantum spin Hall-like effect controlled by multiple-valued
degrees of freedom~\cite{Xu1}, are beneficial to TPs' applications.
Because the topologically protected states are immune to
backscattering~\cite{Kane,Bernevig1757,Fu}, TPs would be very
promising for applications in the abnormal heat transport,
solid-state refrigeration, and phonon wave-guides, and so on.

Unlike the topological electronic materials in which one only needs
to focus on energy states near the Fermi level, phonons exhibit
several distinct properties. First, there are no limits of Pauli
exclusion principle. Second, each phonon mode, following
Bose-Einstein statistics, can become practically active, due to
thermal excitation. Third, phonons are charge neutral and spinless
Bosons, which can not be directly influenced by the electric and
magnetic fields. Hence, a full frequency analysis for all phononic
branches is needed for the study of TPs. To date, a large-scale
identification of TP materials remains challenging, because it is
far more expensive to compute the phonon band dispersions than to
calculate the electronic band structures. Hence, it is certainly
more difficult to seek feasible TP materials in a high-throughput
(HT) computational manner, as compared with the recent works in
topological electronic materials~\cite{ht1,ht2,ht3}. In this work,
we present an efficient and fully automated workflow that can screen
the TP crossings in a large number of solid materials. Our results
reveal that TPs extensively exist in phonon spectra of many known
materials, which can be classified into single Weyl, high degenerate
Weyl, and nodal-line (ring) TPs.

\begin{figure*}
\centering
\includegraphics[width=0.90\textwidth]{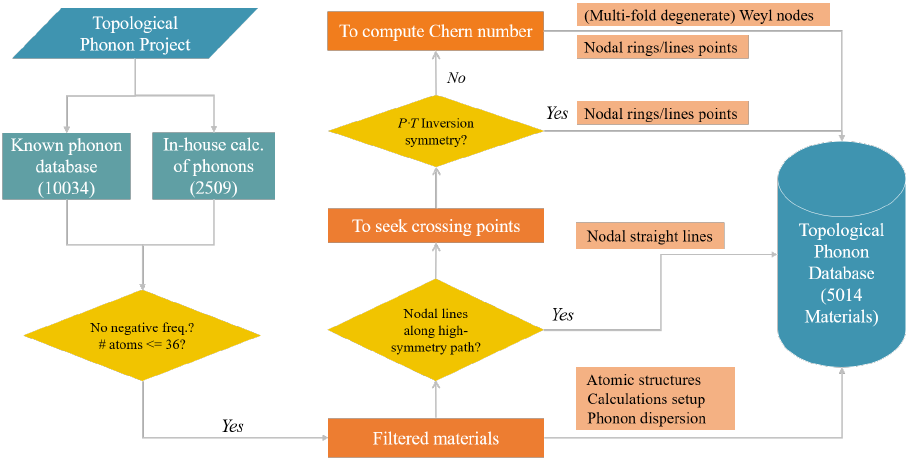}
\caption{The schematic flowchart of high throughput screening on
topological phonons.} \label{fig1}
\end{figure*}

As shown in several prototypical materials \cite{
TPW1,TPW2,TPW2-2,TPW3,TPW4,TPW5,TPW6,TPNL1,TPNL2,TPNL3,TPNL4,TPNL5,
TPHE,Yongxu2019,Niu1,Xu1,Xu2,tp1d,exp1,Yongxu2019}, identifying TPs
requires several stages of manual selection and subjective human
decisions. To enable the TPs discovery in an automatic manner, we
present a high-throughput (HT) screening and data-driven approach to
discover and categorize TPs, as described in Fig.~\ref{fig1}
including the following four steps.

(1) \textit{Phonon data collection}. To obtain phonon spectra for a
large volume of known materials, our approach collects the force
constant data from public phonon database
\cite{phonondb,petretto2018high}, which includes about 10,000
materials. The data set was further augmented by our in-house
computations for over 2000 materials possessing 58 common structural
prototypes. It is well known that the calculated force constants are
numerically sensitive to the choices of several parameters,
\emph{e.g.}, the supercell size, \emph{K}-point mesh and energy
cutoff. To guarantee that the predictions are reliable, we filtered
out the materials with notable imaginary frequencies ($<$ -0.5 THz)
in the whole phononic momentum space.

(2) \textit{Nodal straight lines identification}. We computed their
band dispersions on the automatically generated high symmetry band
paths \cite{HINUMA2017140}. If there exist degenerate phononic bands
along high-symmetry paths, we will compute the Berry phases, by an
integral of Berry connection
($\mathcal{A}_n(q)=i<\mu_{n,q}|\nabla_q|\mu_{n,q}>$) over a closed
$q$ path~\cite{Berry}), for 20 consecutive points on each of these
bands. The bands possessing continuous points with Berry phase
values, $\pm$$\pi$, will be marked as the topologically non-trivial
nodal straight lines.

(3) \textit{Crossing points identification}. For the rest band paths
in the phonon spectra, we systematically scan 50 consecutive points
on each band path. In the entire frequency range, we considered the
points possessing two adjacent eigenfrequencies less than 0.5 THz.
For each point, we performed a minimization based on the conjugate
gradient algorithm to obtain the local minimum of the frequency
difference ($\Delta_{\textrm{freq}}$). The points with
$\Delta_{\textrm{freq}}$ less than 0.2 THz were stored for further
evaluation. After optimization, the identified inversion points may
go anywhere in the entire reciprocal space. Therefore, we also
checked if the points are at or off the high-symmetry paths.

(4) \textit{Crossing points assignment}. The identified phononic
crossing points were then divided into two groups based on the
presence of both inversion ($P$) and time reversal ($T$) symmetries
for each material. In a three-dimensional (3D) system with $PT$
symmetries, the Berry curvatures of non-degenerate phononic bands
are forced to be zero because of the monopole feature of Berry
curvature, and thus, the Weyl TPs would not occur in such system.
Once the phononic bands at a degenerate point has an opposite
non-zero Berry curvature along one direction, such topological
non-trivial degenerate point has to occur continuously by forming
nodal-line (ring) TPs, due to the continuity of phonon wave function
in the 3D momentum space. As a result, we only looked for nodal-line
(ring) TPs in materials with \emph{PT} symmetries.

In noncentrosymmetic materials, the phonon dispersions possibly form
single Weyl or high degenerate Weyl TPs, in addition to nodal-line
(ring) TPs. In order to clarify these three types of TPs, we
introduced another formula of Chern number, which can be derived by
integrating Berry curvatures of a close surface~\cite{Berry}
according to
\begin{math}
n = \frac{1}{2\pi}\int_{\textbf{\emph{S}}}d\textbf{\emph{S}} \cdot
\Omega(\textbf{\emph{q}}).
\end{math}
Here, \textbf{\emph{S}} is a closed surface which wraps the target
crossing point, and the $\Omega(\textbf{\emph{q}})$ is the Berry
curvature at the phonon momentum \textbf{\emph{q}} on the selected
closed surface. For the isolated crossing points at the
high-symmetry band paths, we marked the points with nonzero integer
Chern numbers (\emph{e.g.}, $\pm$1, $\pm$2) as single or high
degenerate Weyl points. Otherwise, they would be labeled as
nodal-ring points, given the fact that the nodal-line points were
already extracted in step (2). Of course, it needs to be emphasized
that many materials may yield many crossing bands along
off-high-symmetry paths, which can be separate Weyl TPs or
nodal-line (ring) TPs. In principles, this approach can be easily
extended to investigate these crossing points.

In total, our approach revealed that 5014 materials exhibit TPs
states (supplementary Table S1). Among them, we have identified
three main categories of single Weyl, and high degenerate Weyl, and
nodal-line (ring) TP materials. Although people have suggested
several possible routes to break the time-reversal symmetry (TRS),
such as, Lorentz force for ionic lattices \cite{Loren} and
spin-lattice interactions for magnetic lattice ~\cite{sl}, it is
impossible to break the TRS for phonon without any tunable external
field. Therefore, we did not attempt to identify the intrinsic
topological phonon insulators in our data. Here, we present three
typical cases for each group, including half-Heusler LiCaAs alloy
(single Weyl TPs), superconducting BeAu (high degenerate Weyl TPs),
and ScZn with the \emph{PT} symmetries (nodal-line/ring TPs).

\begin{figure*}
\centering
\includegraphics[width=0.90\textwidth]{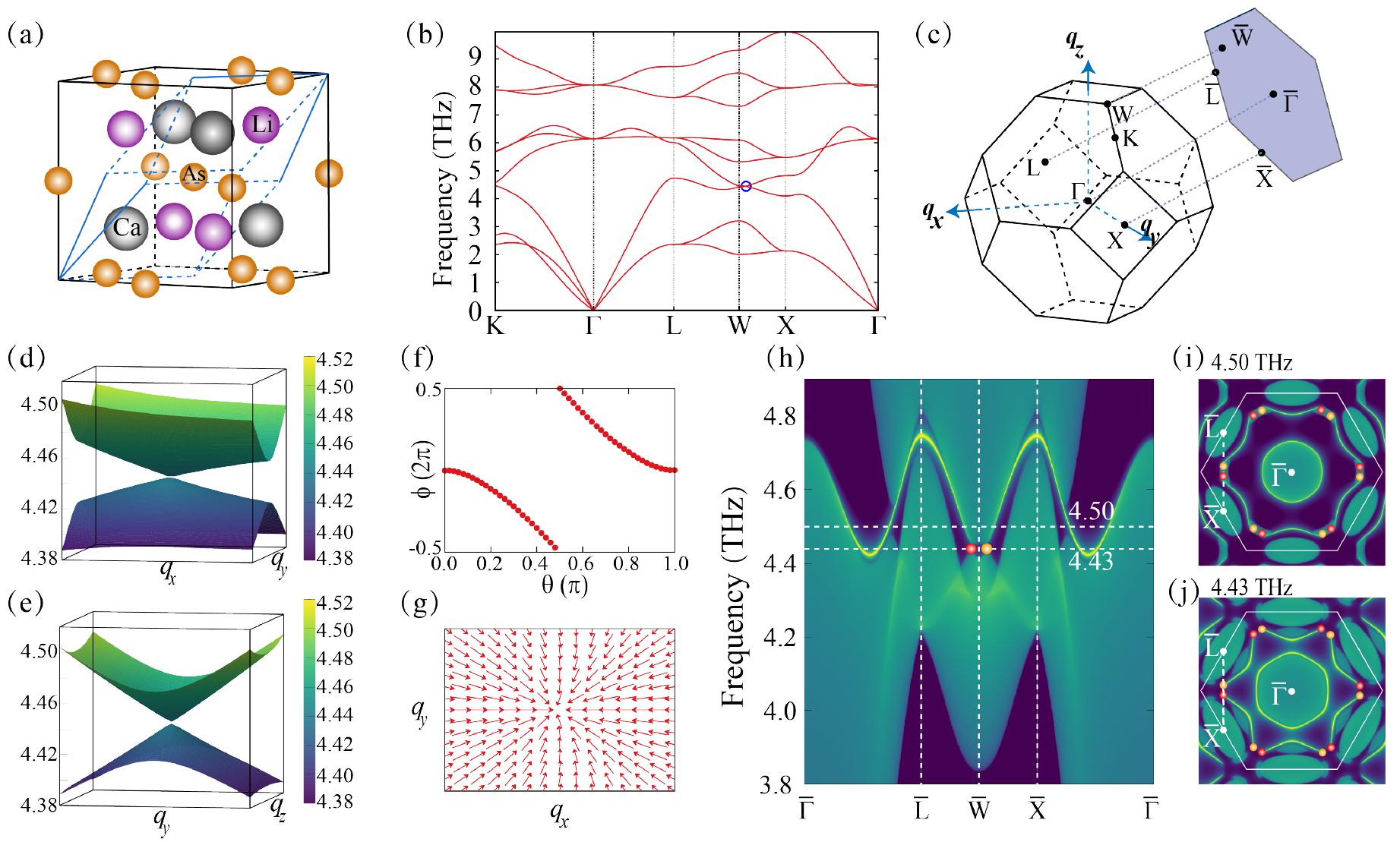}
\caption{\textbf{Phonon dispersion and single Weyl TPs of LiCaAs.}
\textbf{a}, the unit cell and primitive cell of LiCaAs (space group
$F\bar{4}3m$ 216), \textbf{b}, the phonon dispersion along high
symmetry line. The blue circle marks the crossing point of the
single Weyl TPs. \textbf{c}, the bulk BZ and the (111) surface BZ.
\textbf{d} and \textbf{e}, the 3D phonon dispersions centered at the
Weyl point. \textbf{f} and \textbf{g}, the Wannier center evolution
and Berry curvatures distributions around this Weyl phonon.
\textbf{h}, the surface states along high symmetry line. The red and
yellow circles represent the projected Weyl phonon with positive and
negative topological charges of 1 and -1, respectively. \textbf{i}
and \textbf{j}, the Weyl phonon induced non-trivial surface phononic
arc states at 4.50 and 4.43 THz, respectively.} \label{fig2}
\end{figure*}

Half-Heusler compounds $ABX$ ($A$ = Li, Na, K, Rb, Cs; $B$ = Mg, Ca,
Zn; $X$ = P, As) exhibit similar phonon spectra and here we use
LiCaAs as an example (Fig.~\ref{fig2} a). At the frequency of 4.590
THz, the highest longitudinal acoustic and the lowest transverse
optical branches cross at the (0.5000, 0.2175, 0.5175) point along
the \emph{W}-\emph{X} high-symmetry path (Fig~\ref{fig2} b). This
double degenerate point is protected by $C_2$ symmetry at the BZ
boundary. The phononic dispersions around this crossing point holds
the Weyl nodal feature, as shown in the 3D phononic dispersions in
\emph{q}$_{xy}$ and \emph{q}$_{yz}$ in Fig~\ref{fig2} d and
~\ref{fig2}e. To further confirm its topological nature, the Wannier
centers evolution has been derived for the 3rd phononic band
(Fig~\ref{fig2} f). It gives a negative topological charge of -1 and
this crossing acts like the sink of Berry curvature distributions in
Fig~\ref{fig2} g. These results indicate that the crossing along the
$X$-$W$ path is a single type-I Weyl TP. The nontrivial surface
states of the (111) surface along the high-symmetry line have been
derived in Fig~\ref{fig2} h and the opening arcs connecting two
projected Weyl nodes with opposite topological charges can be
observed at 4.43 and 4.50 THz (Fig~\ref{fig2} i-j), respectively.

\begin{figure*}
\centering
\includegraphics[width=0.90\textwidth]{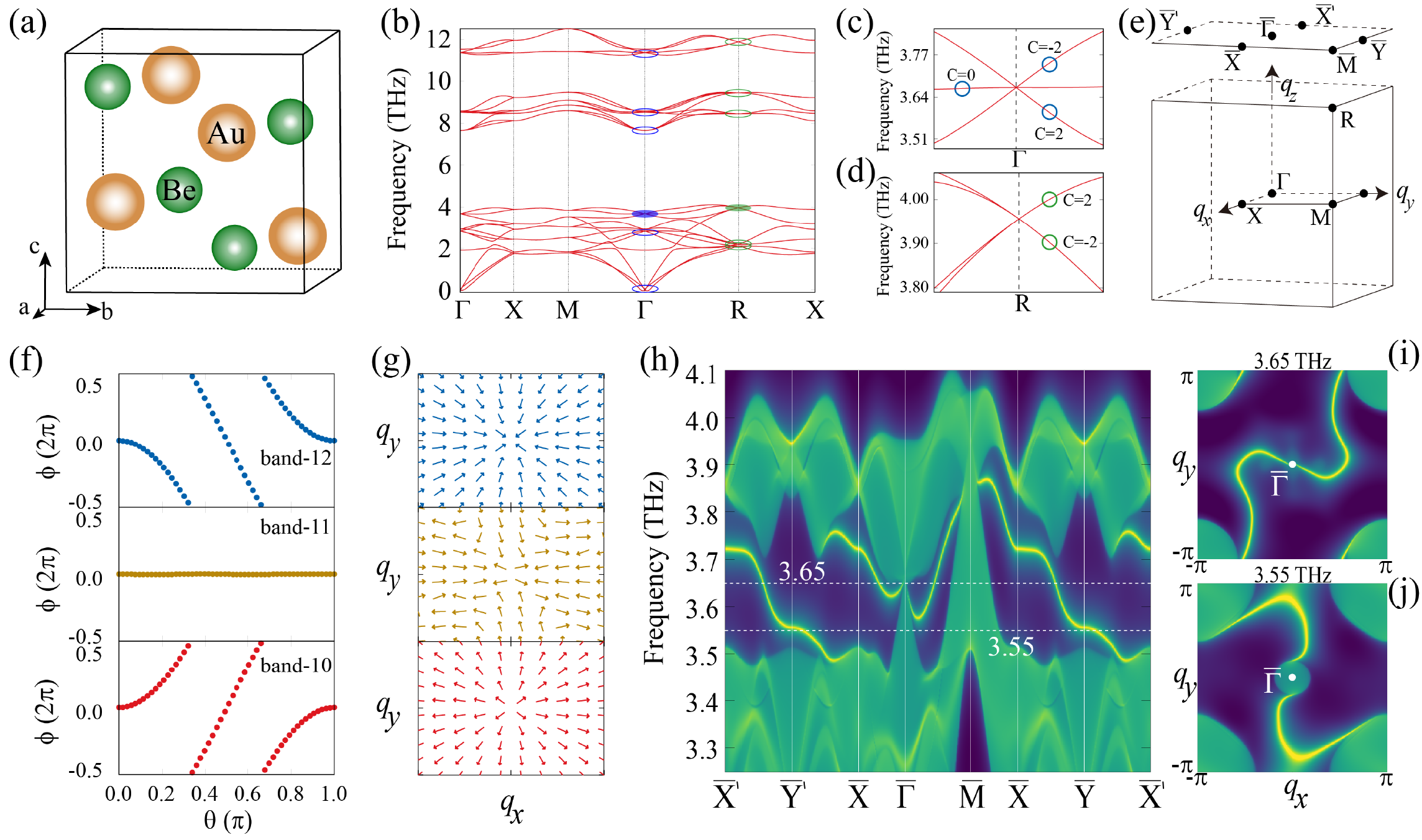}
\caption{\textbf{Phonon band structures and surface states for
topological high degenerate Weyl TPs in BeAu.} \textbf{a}, the
crystal structure of BeAu (space group $P2_13$ 198). \textbf{b},
phonon dispersion along high-symmetry lines. the blue circles are
spin-1 Weyl points at $\Gamma$ and the green circles are charge-2
Dirac points at \emph{R}. \textbf{c}, spin-1 Weyl point of 3.669 THz
at $\Gamma$. \textbf{d}, charge-2 Dirac point of 3.956 THz at
\emph{R}. \textbf{e}, bulk and surface BZ. \textbf{f}, the Wannier
center evolution for three branches 10, 11 and 12 centered at the
$\Gamma$. \textbf{g}, the Berry curvatures distributions of three
branches 10, 11 and 12 at the centered $\Gamma$. \textbf{h}, the
surface local density of states for (001) surface along
high-symmetry directions. \textbf{i} and \textbf{j}, the
corresponding surface arcs at 3.65 and 3.55 THz, respectively. Even
though BeAu exists in reality, we still found that around $\Gamma$
point one acoustic branch of BeAu shows the extremely small
imaginary frequency, which cannot be removed in our current
calculations, possibly due to misconsideration of long range
interatomic interaction in the force constant construction or
anharmonic effects.} \label{fig3}
\end{figure*}

\begin{figure*}
\centering
\includegraphics[width=0.90\textwidth]{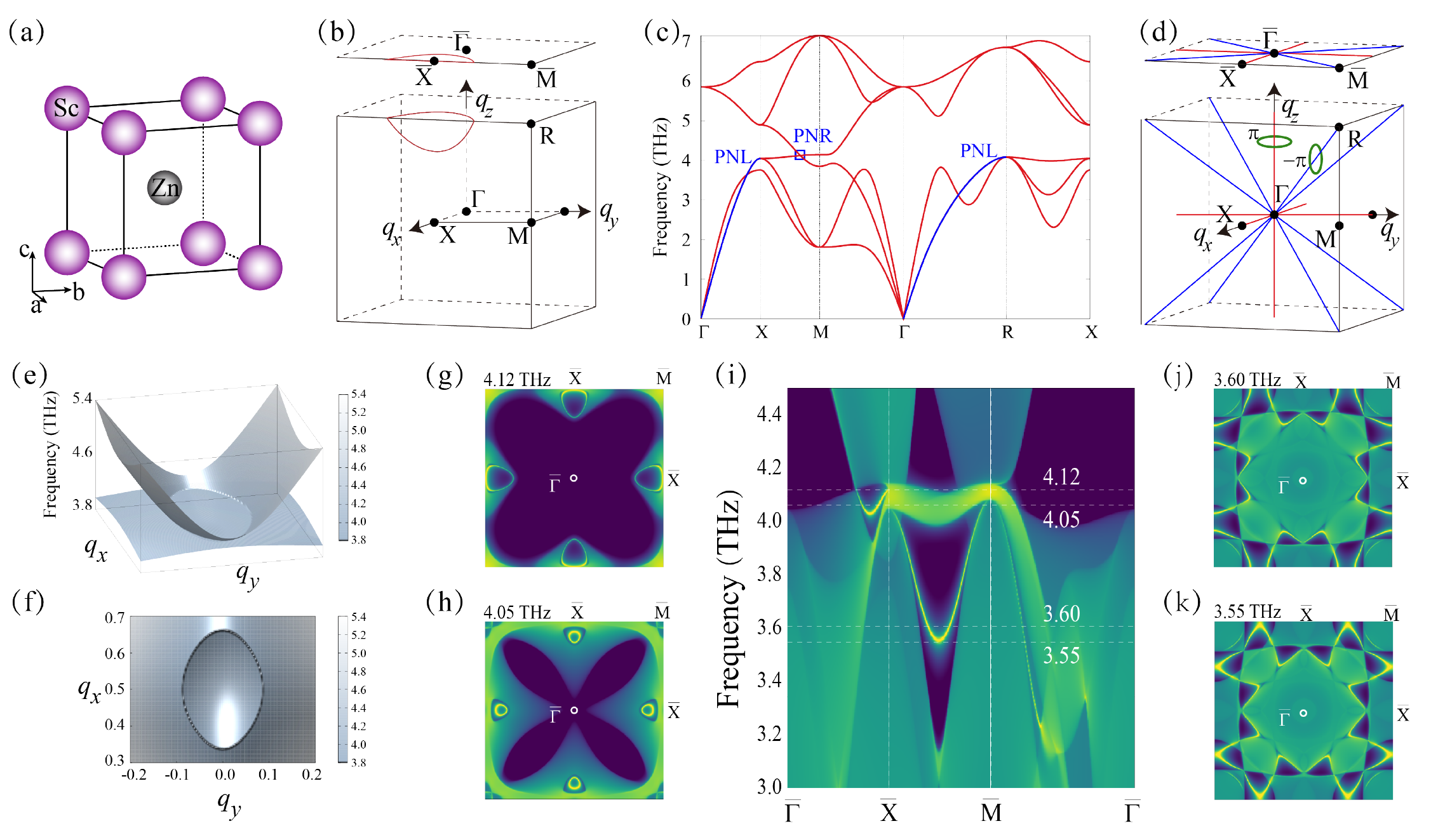}
\caption{\textbf{Phonon band structures and topological properties
of phonon of ScZn.} \textbf{a}, The crystal structure of ScZn (space
group \textit{Pm$\bar{3}$m} 221). \textbf{b}, The BZ of ScZn and the
illustration of nodal rings (red curve). \textbf{c}, The phonon
spectrum of ScZn. \textbf{d}, The illustrations of the nodal
straight lines along the $\Gamma$-\emph{R} and $\Gamma$-\emph{X}
directions in ScZn. \textbf{e} and \textbf{f}, The 3D phonon bands
around the phonon nodal line surrounding the \emph{M} point.
\textbf{g} and \textbf{h}, The derived phononic surface states at
the frequencies of 4.12 THz and 4.05 THz, respectively, of the (001)
surface BZ as defined in panel \textbf{b}. \textbf{i}, The phononic
surface spectrum of the (001) surface. \textbf{j} and \textbf{k},
The phononic surface states at the frequencies of 3.60 THz and 3.55
THz, respectively.} \label{fig4}
\end{figure*}

BeAu is a typical non-centrosymmetric superconductor with a
transition temperature of 3.2 K~\cite{BeAu1} and its structure
crystallizes in a cubic symmetry with a space group of No. 198
($P2_13$, Fig.~\ref{fig3} a). In it phonon dispersion
(Fig.~\ref{fig3} b), there exists two kinds of high degenerate Weyl
points. At the $\Gamma$ point, the Hamiltonian of six threefold
degenerate Weyl TPs (see blue circles in Fig.~\ref{fig3} b and c)
and can be written as $\mathbf{H_3(k) \varpropto q \cdot C}$, where
$\mathbf{q}$ is a wavevector and $\mathbf{C_i}$ are the rotation
generators for spin-1 bosons ~\cite{TPW1}. As shown in
Fig.~\ref{fig3}c, each high degenerate Weyl node has three
degenerate bands with respective Chern numbers of +2, 0 and -2. As
illustrated in Fig.~\ref{fig3}(f-h), the Wannier center evaluations
for three phononic branches of the $\Gamma$ point with the frequency
of 3.669 THz confirm that their nontrivial topological charges and
their corresponding Berry curvatures show the sink and source
behaviors. At the \emph{R} point, the phonon spectrum exhibits six
nontrivial fourfold degenerate Weyl points in Fig.~\ref{fig3} b and
d. For each Weyl point, the Hamiltonian can be written as
$\mathbf{H_4 (k) \thicksim I_2 \bigotimes (q \cdot \sigma)}$ where
$\mathbf{I_2}$ are the $2$-order identity matrix and
$\mathbf{\sigma_i}(i = x, y, z)$ are the three Pauli matrices. These
high degenerate Weyl nodes induce the topological non-trivial
surface states in Fig.~\ref{fig3} h, indicating that the two long
non-trivial arcs connect two projected high degenerate Weyl points
at $\bar{\Gamma}$ and $\bar{M}$.

ScZn (Fig.~\ref{fig4}a) is a well-known Zintle compound
\textit{MX}~\cite{ScZn} (\emph{M} = Sc, Y, La; \emph{X} = Zn, Cd,
Hg). In its phononic dispersion, there exist six nodal-ring TPs
(Fig.~\ref{fig4}e and \ref{fig4}f), surrounding the $M$ point on the
BZ boundary (Fig.~\ref{fig4} b). ScZn also host phononic nodal
straight lines along both $\Gamma$-X and the $\Gamma$-R directions
(Fig.~\ref{fig4}c and d). Both the nodal-ring and nodal straight
line TPs exhibit the topological nontrivial feature with nonzero
Berry phases. Their occurrences are protected by the mirror symmetry
of ScZn. We have further derived the nontrivial surface TPs in
Fig.~\ref{fig4} i. Two types of the drumhead-like nontrivial surface
TP states can be observed, one is the nodal-ring TP induced closed
loops around the $\bar{X}$ point (Fig.~\ref{fig4} g-h) and the other
is the straight nodal line induced special pattern with fourfold
symmetry on the (001) surface (see Fig.~\ref{fig4}(j-k)).

Furthermore, in order to be experimentally observable, a candidate
material is expected to possess distinct Weyl or nodal-line (ring)
TPs and possible clean nontrivial surface TPs. For this purpose, we
mathematically define the clean TP states for the nontrivial
crossing points satisfying two conditions in bulk phonon spectra,
(\emph{i}) the crossing points have to be located at local minima
with negligible or zero phononic density of states (DOS) ($<$ 0.01
states/atom/THz) and (\emph{ii}) the dispersion at the local minima
is sufficiently large ($\partial E/\partial q > $ 3 THz$\cdot$\AA).
On basis of these two criteria, we have filtered 322 clean TP
materials (supplementary Table S5). For instance, The single type-I
Weyl TPs in LiCaAs are clean, because they do not overlap with the
other bulk phonon branches and have a zero phonon density at the
frequency of 4.590 THz (Fig.~\ref{fig2} i and j). Among those 5014
TP materials 34 materials only have single Weyl TPs (supplementary
Table S2) and 463 materials host mixed single Weyl TPs and
nodal-line (ring) TPs as well as 268 materials host mixed single
Weyl, high degenerate Weyl and nodal-line (ring) TPs (supplementary
Table S4). In addition, we have found 449 materials host high
degenerate Weyl TPs ( supplementary Table S4) and 4977 materials for
topological nodal-line (ring) TPs (in supplementary Tables S3, S4
and S5).

In summary, we have developed a HT and data-driven approach to
evaluate TPs over 10,000 materials using the existing
phononic database and our in-house calculations. Our screening
suggest that TP states are universally present in many materials,
highlighting extensive possibilities for investigation of TPs and
their potential applications in many fields.



\subsection*{COMPUTATIONAL METHODOLOGIES}
All DFT calculations have been performed by Vienna \textit{ab
initio} simulation package (VASP), based on the projector augmented
wave (PAW) potentials and the generalized gradient
approximation(GGA) within the Perdw-Burke-Ernzerhof (PBE) for the
exchange correlation treatment. The force constants downloaded from
the PHONONPY database were calculated by the finite displacement
method. For the in-house phonon calculations, we used the density
function perturbation theory (DFPT). We performed the geometry
optimization of the lattice constants by minimizing the forces
within 0.001 eV/{\AA}. The cut-off energy for the expansion of the
wave function into the plane waves was set to 1.5 times of the ENMAX
in the POTCAR. For the topological analysis, we used the conjugate
gradient method in SciPy \cite{scipy} to get the crossings and
calculated the Berry phase and Chern number to identify the
nontrivial topological natures. To determine the topological charge,
the Wilson-loop method~\cite{wl1,wl2} was chosen. A sphere centered
at a WP was sliced into independent orbitals by a constant polar
angle $\theta$ and the evolution of Wannier centers (phase factor
$\phi$) on orbitals can give topological charges of WPs. For the
surface DOS, we used force constants as tight-binding parameters to
construct surface and bulk Green's functions and the imaginary part
of the Green's function produces the DOS ~\cite{Green1,Green2}.

\subsection*{SUPPLEMENTAL INFORMATION}
We have provided a supplementary material including 14,246 pages and
5014 figures to classify all 5014 TP materials according to the
geometrical character and the Berry phases of topological nodal
points (Weyl node, Dirac node, and high degenerate nodal points) and
nodal line (ring) TPs. Each material entry includes the spatial
information of the points (\emph{e.g.}, $x, y, z$ coordinates,
frequencies, modes and band paths), and multiplicity, degeneracy,
and topological charges to each phononic band crossing points. These
data, together with the interactive visualization of atomic
structures and phonon band dispersion, are also available, if
requested.

In order to effectively and conveniently analyze topology of
phonons, we developed an \texttt{HT-TPHONON} code to automate all
processes (as shown in Fig.~\ref{fig1}) and connect them with the
DFT calculations based on Python scripting. The code will be
released upon the publication of this manuscript.

\section*{ACKNOWLEDGEMENTS}
Work at IMR was supported by the National Science Fund for
Distinguished Young Scholars (No. 51725103) and by the National
Natural Science Foundation of China (Grant No. 51671193). Work at
UNLV is supported by QZ's startup grant. All calculations have been
performed on the high-performance computational cluster in Shenyang
National Park and XSEDE (TG-DMR180040).

\section*{AUTHOR CONTRIBUTIONS}
X.-Q.C first proposed this idea and both X.-Q.C and Q. Z. designed
the research. J.X.L, J.X.L, R.H.L, L.W, Q.Z., and X.-Q.C performed
and analyzed the calculations and contributed to interpretation and
discussion of the data. S.A.B and Q.Z coded the online topological
phonon database with the help of J.X.L and X.-Q.C. X.Q.C, J.X.L and
Q.Z. wrote the manuscript. All authors discussed this manuscripts.
\bibliography{ref}

\end{document}